\documentclass[pdflatex,sn-mathphys-ay]{sn-jnl}

\usepackage{graphicx}%
\usepackage{multirow}%
\usepackage{amsmath,amssymb,amsfonts}%
\usepackage{amsthm}%
\usepackage{mathrsfs}%
\usepackage[title]{appendix}%
\usepackage{xcolor}%
\usepackage{textcomp}%
\usepackage{manyfoot}%
\usepackage{booktabs}%
\usepackage{algorithm}%
\usepackage{algorithmicx}%
\usepackage{algpseudocode}%
\usepackage{listings}%

\newcommand{\Lagr}{\mathcal{L}}

\begin{document}

\title[Article Title]{SENSE - \textbf{S}ensor-\textbf{E}nhanced \textbf{N}eural \textbf{S}hear Stress \textbf{E}stimation for Quantitative Oilfilm Visualizations}


\author*[1]{\fnm{Lennart} \sur{Rohlfs}}\email{l.rohlfs@tu-berlin.de}

\author[1]{\fnm{Julien} \sur{Weiss}}

\affil*[1]{\orgdiv{Chair of Aerodynamics}, \orgname{TU Berlin}, \orgaddress{\street{Marchstr. 12-14}, \city{Berlin}, \postcode{10587}, \country{Germany}}}


\abstract{Wall shear stress ($\tau_w$) quantification is fundamental in fluid dynamics but remains challenging in wind-tunnel experiments. Sensor-based methods offer high accuracy but lack spatial resolution for capturing complex three-dimensional effects. Conversely, oil-film visualization is a simple method to obtain high-resolution surface flow topology by processing a sequence of images using optical flow (OF) techniques. However, leveraging this approach for quantitative analysis suffers from noise and systematic biases. This study introduces SENSE (Sensor-Enhanced Neural Shear Stress Estimation), a data-driven approach that leverages a neural network to enhance OF-based $\tau_w$ estimation through the integration of sparse, high-fidelity sensor measurements via a multi-objective loss function. SENSE processes oil-film image sequences directly, inherently mitigating temporal noise without explicit averaging. The method is validated in a turbulent separated flow on a one-sided diffuser. Results demonstrate SENSE's robustness to sequence length and spatial resolution compared to classical optical flow algorithms. Crucially, incorporating sparse sensor data significantly improves quantitative accuracy, achieving over 30\% reduction in root-mean-squared error on validation sensors with only 8 strategically distributed sensors. The sensor data provides a global regularization effect, improving estimates far from sensor locations. SENSE offers a promising approach to elevate oil-film visualization to a reliable quantitative measurement technique by combining image sequences and sparse sensor data.}




\maketitle

\section{Introduction}\label{sec:introduction}

The characterization of wall shear stress ($\tau_w$) is fundamental to understanding and predicting fluid dynamic phenomena, including boundary layer development, flow separation, transition, and skin friction drag. Unfortunately, wall shear stress is also a very challenging quantity to accurately quantify in a wind-tunnel environment. This has led to a large amount of techniques being developed over the years that are summarized in review articles by \cite{Winter1979}, \cite{Haritonidis1989}, \cite{Naughton2002} and more recently \cite{Orlu2020}. To give a brief overview, techniques to measure wall shear stress can be broadly categorized into direct and indirect methods. The direct methods include floating element sensors that employ some mechanical structure that is deflected by the near-wall flow, and film-based techniques such as oil-film interferometry \citep{Tanner1976} and liquid-crystal coatings \citep{Reda1994} where the shear stress is inferred from the deformation patterns in the surface coatings after an extensive calibration. Alternatively, indirect methods measure another quantity such as pressure or heat transfer and link it to the shear stress through a known relation. Examples include Preston- or Stanton tubes \citep{Head1962, Trilling1955} that are used to determine the local mean shear stress from pressure measurements at the wall. If the fluctuating shear stress is also required, thermoelectric methods such as surface hot-wires and microelectromechanical systems (MEMS)-based sensors have been used to great success \citep{Sheplak2004, Weiss2024}.\\
A common drawback of the sensor-based methods is that they only provide the shear stress at a single location. Furthermode, the use of dozens or more sensors to increase the spatial resolution is often impractical due to space, time, or budget constraints. For this reason, researchers often supplement sparse shear stress measurements with surface oil-film visualizations that offer a simple way of obtaining a qualitative map of the shear stress field. They are typically performed by coating the model with a thin layer of an oil-pigment mixture that is deformed into streaky patterns during a wind-tunnel run \citep{Merzkirch1987}. After the experiment, a picture is taken and skin friction lines are sketched on the resulting image revealing the near-wall topology (e.g, \cite{Ruderich1986}). This process is traditionally performed by hand and thus relatively inefficient and inherently subjective. To automate this process, recent advancements in the field of deep learning techniques include training convolutional neural networks (CNNs) to relate oil-flow texture to the shear stress direction \citep{Schulte2025}. Alternatively, applying optical flow (OF) techniques to image sequences also enables the extraction of skin friction lines as demonstrated by \cite{Liu2008b} and \cite{Rohlfs2024a}. Leveraging this approach for quantitative shear stress estimation, however, presents further challenges.\\
Mathematically, the OF method is based on the relationship between the thickness and the luminescent intensity of the recorded oil-film that was derived from the thin oil-film equation by \cite{Liu2008}. This equation has the same form as the optical flow equation with the exception of an additional term that models the effect of the pressure gradient and gravity. Although neglecting this term may suffice for determining the qualitative skin friction topology \citep{Liu2013}, it introduces a systematic bias that limits the accurate estimation of the quantitative shear stress. Furthermore, the OF estimations from individual images are typically noisy and contain unsteady effects that are inherent in turbulent flows. Therefore, a sequence of solutions is averaged which introduces further inaccuracies due to transport, accumulation, or evaporation of the oil-film over time that violates the brightness constancy assumption of the optical flow equation \citep{Horn1981DeterminingFlow}. This has motivated the development of methods capable of processing sequences without explicit averaging such as the linear least-squares approach by \cite{Lee2018}, but the method has not yet been used with experimental data and its robustness towards noise and optical artifacts is questionable.\\
To bridge the gap between sparse, accurate point measurements and high-resolution, qualitative field visualizations, we propose a data-driven approach called \textbf{SENSE} (\textbf{S}ensor-\textbf{E}nhanced \textbf{N}eural \textbf{S}hear stress \textbf{E}stimation), which aims to improve the optical flow estimation with sparse sensor measurements. The method utilizes a neural network trained on an arbitrary number of frames, inherently handling temporal variations and noise without explicit averaging. The multi-objective loss function is composed of a data term derived from the linearized OF equation, a regularization term for smoothness and a sensor term that anchors the network output with sparse but accurate sensor measurements, mitigating systematic biases from unmodeled physics or oil-film property variations. This approach of embedding physical constraints and data points into the loss function shares similarities with the framework of Physics-Informed Neural Networks (PINNs) that have been applied recently to many fluid mechanical problems \citep{Karniadakis2021, Rohlfs2024}. The primary objective of this study is to systematically evaluate how effectively SENSE, through its multi-frame processing and sensor integration, overcomes the previously discussed limitations compared to classical OF algorithms in a real-world experimental context. To determine the accuracy of the predictions, we acquire a database of reference shear stress measurements in a turbulent, separated flow over a one-sided diffuser. The flow is characterized by a complex, fully three-dimensional flow topology due to corner effects and secondary flows that are typical of turbulent separation bubbles generated in rectangular test sections \citep{Steinfurth2024}.\\
The remainder of this paper is structured as follows: Section \ref{sec:exp_setup} details the experimental setup, including the wind tunnel facility, the oil-film visualization, and the MEMS reference sensors. Section \ref{sec:methodology} provides a background on optical flow analysis and introduces the proposed SENSE algorithm. Section \ref{sec:results} presents the comparative results and analyzes the performance enhancements achieved through sensor integration. Finally, Section \ref{sec:conclusion} summarizes the key findings and discusses potential options for future work.

\section{Experimental Setup}
\label{sec:exp_setup}
\begin{figure}[htbp]
\centering
\includegraphics[width=0.45\textwidth]{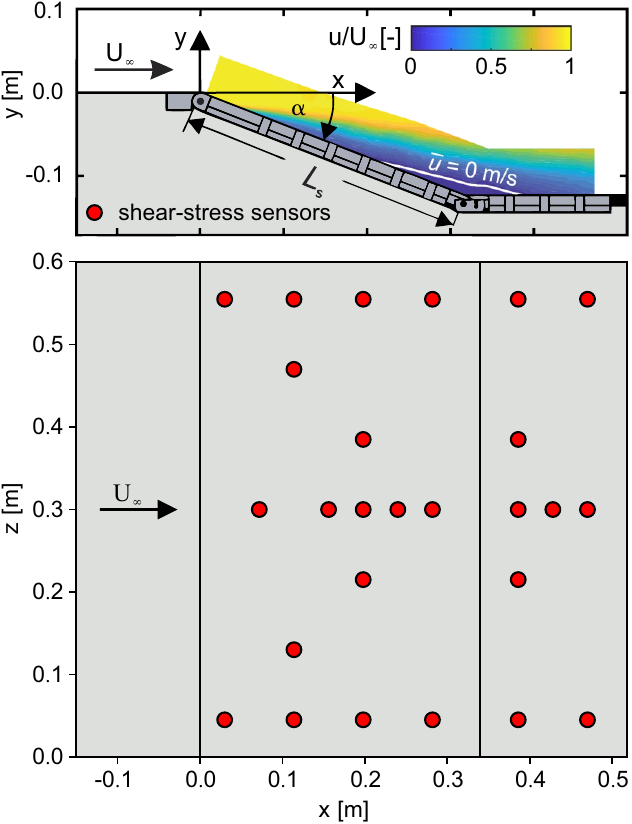}
\caption{Geometry of the one-sided diffuser; Side view with contour of the longitudinal velocity field on test-section centerline (top); Top view with shear-stress sensor locations (bottom)}\label{fig:MEMS_setup}
\end{figure}
The experiments are conducted in a temperature-regulated, closed-loop wind tunnel at a nominal velocity $U_{ref} = 20$ m/s. Side and top views of the test-section geometry are shown in Fig. \ref{fig:MEMS_setup}. It has a width of $0.6$ m, a backward-facing ramp on the bottom and a flat ceiling $0.4$ m above the ramp corner \citep{Hecklau2013}. For this study the ramp with a length of $L_s=0.34$~m is set to an inclination angle of $20^\circ$. The resulting adverse pressure gradient causes a pressure-induced turbulent separation bubble to form at the foot of the ramp as highlighted by the contour plot of the longitudinal velocity field at the test-section centerline (Fig. \ref{fig:MEMS_setup}). The boundary layer upstream of the diffuser is naturally developed and fully turbulent with a Reynolds number based on momentum thickness of $Re_\theta \approx 1000$, while the Reynolds number based on the ramp length is approximately $350,000$. The mean topology and dynamic behavior in this flow configuration are detailed in \cite{Steinfurth2024} and \cite{Weiss2022}, respectively.

\subsection{Oil Film Visualization}
\begin{figure}[htbp]
\centering
\includegraphics[width=0.45\textwidth]{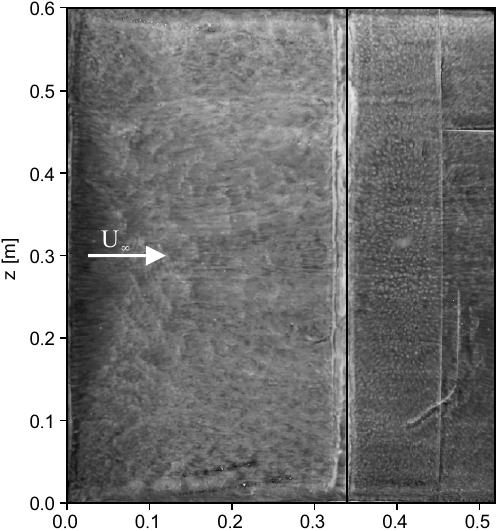}
\caption{Example for a pre-processed frame of the oil-film visualization}\label{fig:oilfilm}
\end{figure}

Surface oil-film visualization was performed using a mixture of flaxseed oil and turpentine, supplemented with green fluorescent particles. This specific mixture was chosen to achieve an oil film that is only weakly affected by gravity while remaining highly sensitive to the flow. UV lighting was used to illuminate the fluorescent particles from both sides of the test section. The image acquisition was performed using a high-resolution DSLR camera mounted perpendicular to the diffuser surface. The camera settings were adjusted to capture two images per second with at a resolution of 8 Mpx resulting in a spatial resolution of $\approx 0.25$ mm/px. Sequences of typically 20-30 seconds (40-60 frames) were recorded. The acquired images are then projected on a wall-bound coordinate system using a homography matrix to account for the camera perspective and the ramp geometry. During this step the spatial resolution can also be modified by adapting the size of the projection grid. An example for such a pre-processed frame can be seen in Fig. \ref{fig:oilfilm}.
\subsection{MEMS Reference Measurements}
Quantitative reference measurements of the wall shear stress vector, $\vec{\tau}_w$, are acquired using calorimetric Micro-Electro-Mechanical Systems (MEMS) sensors. These sensors are based on the design introduced by \cite{Weiss2017a}. Each sensor consists of three slender beams, typically platinum-plated silicon nitride (SiN$_x$), suspended over a micromachined cavity etched into the silicon substrate (see Figs. 3 and 5 in \cite{Weiss2024}). The central beam acts as a heater, maintained at a constant overheat above the ambient fluid temperature using a Constant-Temperature Anemometer (CTA) circuit. The two flanking beams function as detectors, sensing the temperature distribution in the heater's wake. In the presence of near-wall airflow, convective cooling creates an asymmetric temperature profile around the heater that is detected as a temperature difference, $\Delta T_{det}$, between the upstream and downstream detectors via a Wheatstone bridge circuit. A key advantage of this calorimetric principle is its inherent directional sensitivity: the sign and magnitude of $\Delta T_{det}$ (and thus the sensor's output voltage, $E_{det}$) correlate directly with the direction and magnitude of the wall shear stress component along the sensor's axis. The sensors typically exhibit a measurement range on the order of $\pm 10 \, \text{Pa}$ and a flat frequency response up to approximately 1 kHz with a reported uncertainty of $\Delta\tau_w/\tau_w \sim -15\%/+5\%$ \citep{Weiss2024}. Static calibration is performed in a controlled turbulent channel-flow experiment in which the pressure gradient is measured and linked to the wall shear stress \citep{Weiss2017a}.\\
In the present diffuser experiment, an array of 26 such sensors is distributed across the measurement surface, as indicated in Fig. \ref{fig:MEMS_setup}. Similar to the results of \cite{Steinfurth2024}, the streamwise and spanwise wall shear stress components were acquired sequentially at each location by rotating the sensor $90^\circ$ between measurements. To capture the spanwise distribution in the symmetry plane, 8 sensors are aligned along the centerline ($z=0.3 \, \text{m}$). Additional spanwise locations are instrumented with 6 sensors each near the sidewalls ($z=0.045 \, \text{m}$ and $z=0.555 \, \text{m}$), and another 6 sensors are positioned at intermediate spanwise coordinates to provide further reference points. Data is acquired at $4$ kHz for 10 seconds using a 16 bit National Instruments USB-6363 A/D acquisition card. 
\section{Methodology}
\label{sec:methodology}
During an oil-film visualization, the oil constantly moves over the model surface following the local shear stress vector. Over time, more reflective oil will accumulate in areas with low shear stress, resulting in a higher luminescent intensity in the recordings and vice versa. Thus, by analyzing the change in luminescent intensity over a series of images as an optical flow problem, skin friction lines can be extracted automatically \citep{Liu2008b, Rohlfs2024a}. In this section, the fundamentals of the optical flow technique are introduced as well as the common strategies for solving the equation. Additionally, we discuss caveats when attempting to obtain quantitative information from the computed optical flow field and introduce our proposed neural network based approach for solving the optical flow equation that allows a seamless integration of reference sensor measurements as physical constraints for the solution.

\subsection{Optical Flow Analysis}
The estimation of optical flow is widely used in the field of computer vision and begins with the fundamental assumption of brightness constancy, which states that the intensity value of a point in an image remains consistent over small temporal intervals. This assumption can be expressed through a transport equation:
\begin{align}
    \frac{\partial I(\mathbf{x},t)}{\partial t} + \mathbf{u}(\mathbf{x},t)\cdot \Vec{\nabla}I(\mathbf{x},t) = 0
    \label{eq:gen_OF}
\end{align}
Here $I(\mathbf{x},t)$ is the image intensity at spatial coordinates $\mathbf{x}=(x,y)$ and time $t$ and $\mathbf{u}=(u,v)$ represents the velocity vector (also called optical flow) between subsequent frames. The physical interpretation of this vector in an oil-film visualization directly relates to the local surface flow direction and magnitude, providing a measure proportional to the wall shear stress \citep{Liu2008}. Assuming that the time between consecutive frames $(I_0,I_1)$ is constant, as in a video or time-lapse recording, Eq.\ref{eq:gen_OF} can be simplified to the displaced frame difference equation (DFD):
\begin{align}
    I_0(\mathbf{x}) - I_1(\mathbf{x}+\mathbf{u}(\mathbf{x},t)) = 0.
    \label{eq:DFD}
\end{align}
For small displacements, which can be controlled through appropriate camera settings and sampling rates, the optical flow equation (\ref{eq:gen_OF}) can be linearized using a first-order Taylor series expansion around $(\mathbf{x},t)$, yielding:
\begin{align}
    I_x u + I_y v + I_t = 0
    \label{eq:lin_OF}
\end{align}
where $I_x$ and $I_y$ are the spatial gradients of the first frame and $I_t$ is the temporal intensity difference between consecutive frames. This linearized equation forms the foundation for most optical flow algorithms, including those employed in our work.

\subsection{Solving the OF equation}
Solving Eq. \ref{eq:lin_OF} for $\mathbf{u}$ is an underdetermined problem — often referred to as the aperture problem in computer vision — that requires additional constraints to determine a unique solution. Since its introduction by \cite{Horn1981DeterminingFlow}, many approaches have been proposed to solve the optical flow problem. Here, we describe some classical algorithms with a focus on readily available algorithms as part of the OpenCV open-source computer vision library (v4.9).\\
Many optical flow algorithms reformulate the problem as an optimization task that minimizes a functional $E(\mathbf{u})$ consisting of a data term $J_D$ incorporating Eq.~\ref{eq:lin_OF} and a weighted regularization term $J_R$ that penalizes specific gradient properties of the vector field:
\begin{align}
    E(\mathbf{u}) = J_D(I_{n}, I_{n+1}, \mathbf{u}) + \lambda J_R(\mathbf{u})
    \label{eq:opti_OF}
\end{align}

In their seminal work, Horn and Schunck employed the sum of the first order gradients of $\mathbf{u}$ in the image plane $\Omega$ as the regularization term:
\begin{align}
    J_R = \int_\Omega \lvert \nabla u \rvert^2 + \lvert \nabla v \rvert^2
\end{align}
Although this approach ensures a smooth and differentiable velocity field, it does not allow for discontinuities that naturally occur at edges and boundaries. Additionally, this regularization tends to dampen the magnitude of the entire gradient field \citep{Schmidt2021}, potentially underestimating the true wall shear stress in regions of high gradient.
These issues can be overcome by using different regularization terms such as the sum of the total variation of $\mathbf{u}$ which leads to the formulation of $J_R$ that is used in the TV-L1 algorithm \citep{Zach2007}:
\begin{align}
    J_R = \int_\Omega \lvert \nabla u \rvert + \lvert \nabla v \rvert
\end{align}
If prior information about the velocity field are known, such as divergence-free conditions (incompressibility) for PIV image sequences, the regularization term can also be used to incorporate those physical constraints \citep{Corpetti2006, Schmidt2021a}.\\
\cite{Farneback2003Two-FrameExpansion} proposed another popular algorithm that solves the optical flow problem through polynomial expansion rather than variational methods. For each image pair, the intensity field of the first frame is approximated by a quadratic polynomial:
\begin{align}
    I_n(x,y) \approx a_0 + a_1x + a_2y + a_3x^2 + a_4xy + a_5y^2 
\end{align}
where $a_n$ are the polynomial coefficients. The shifted intensity field $I_{n+1}(x+u\Delta t,y+v\Delta t)$ is then approximated by the same polynomial with shifted coordinates.

To handle larger displacements and improve robustness, the Farnebäck algorithm implements a multi-scale pyramid approach, constructing a hierarchy where each level represents a downscaled version of the previous one. At each pyramid level, the algorithm iteratively refines the flow field by minimizing the difference between polynomial approximations of consecutive frames. This multi-scale strategy makes the Farnebäck method computationally efficient, though potentially more susceptible to noise compared to variational methods—an important consideration in experimental settings with variable lighting conditions.\\
The Dense Inverse Search (DIS) algorithm, introduced more recently by \cite{Kroeger2016}, extends the efficiency advantages of multi-scale approaches while incorporating patch-based matching techniques traditionally employed in sparse optical flow methods like \cite{LucasKanade1981}. By adopting an inverse compositional image alignment approach, DIS circumvents the computational overhead typically associated with dense patch matching, making it particularly promising for real-time applications and large datasets.

\subsection{Sensor-enhanced oil flow reconstruction}
\label{sec:SENSE}
While the previously introduced algorithms perform well for computer vision benchmarks, they are all designed with the assumption of perfectly uniform lighting and no surface reflections, which are conditions that cannot be guaranteed in an experimental oil-film visualization. Additionally, they do not account for variations of the oil film thickness due to pressure gradient effects or gravity. The impact of these physical properties can be mathematically expressed as an additional term on the right-hand side of the linearized optical flow equation (Eq.~\ref{eq:lin_OF}), as demonstrated by \cite{Liu2008}. However, in typical wind tunnel experiments, neither the pressure gradient nor the exact properties of the oil-film are precisely known, preventing accurate estimation of this additional term. Furthermore, the oil-film usually does not evolve homogeneously over time which causes the optical flow estimations from individual image pairs to be noisy and potentially inaccurate. While this averaging improves the smoothness of the vector field, it can adversely affect the magnitude estimation, as the oil film may begin to dry or be carried away over the duration of a recording, introducing systematic temporal biases.\\
Both previously described effects contribute to a systematic error that may not be negligible when attempting to calibrate the magnitude of the optical flow field with in-situ reference measurements of the shear stress. To circumvent these challenges, we propose a novel data-driven approach called SENSE (\textbf{S}ensor-\textbf{E}nhanced \textbf{N}eural \textbf{S}hear stress \textbf{E}stimation) that directly incorporates sparse shear stress measurements to anchor the optical flow prediction to known values at specific spatial coordinates. Our approach expands upon the optimization functional presented in Eq.~\ref{eq:opti_OF} by introducing an additional term $J_S$:
\begin{align}
    E(\mathbf{u}) = J_D(I_{n}, I_{n+1}, \mathbf{u}) + \lambda_1 J_R(\mathbf{u}) + \lambda_2 J_S(\mathbf{u}, \tau_w)
    \label{eq:sensor_OF}
\end{align}
This formulation introduces two weighting parameters, $\lambda_1$ and $\lambda_2$, where the additional term $J_S$ establishes a relationship between the reference shear stress $\tau_w$ from an arbitrary number of sensor measurements and the optical flow $\mathbf{u}$ at corresponding positions. To ensure this comparison is made on a consistent scale, we normalize both quantities with their respective maximum values (denoted as $\overline{\mathbf{u}}$ and $\overline{\tau}_w$). The sensor regularization term is then defined as:
\begin{align} J_S = \sum_{\mathbf{x}_i \in S} ||\overline{\mathbf{u}}(\mathbf{x}_i) - \overline{\tau}_{w}(\mathbf{x}_i)||^2 \label{eq:Js_vector_sum_sq_norm} \end{align}
\begin{figure}[htbp]
\centering
\includegraphics[width=0.49\textwidth]{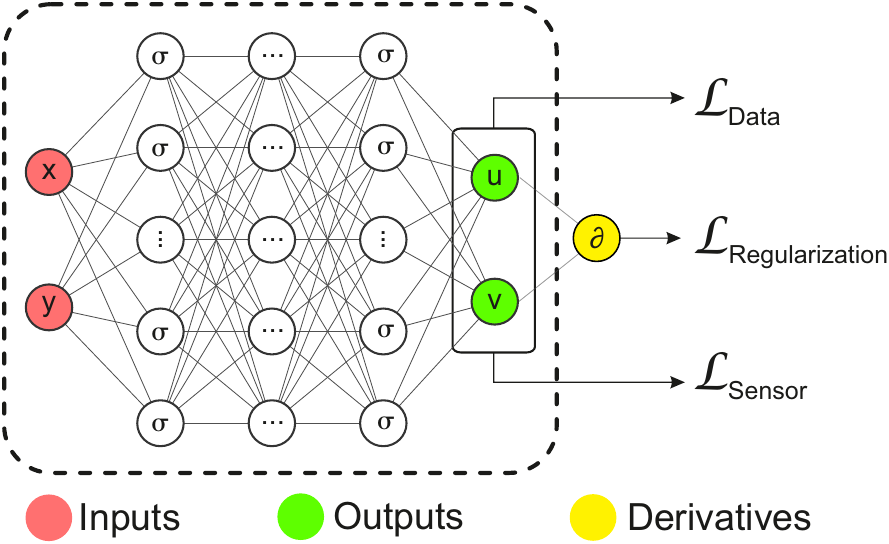}
\caption{Schematic of the neural network architecture. The 2D coordinates $x,y$ are mapped to the optical flow components $u,v$ through a number of hidden layers.}\label{fig:NN_arch}
\end{figure}
To find the vector $\mathbf{u}$ that minimizes $E(\mathbf{u})$ we use a Multi-Layer-Perceptron (MLP) neural network that maps spatial coordinates $(x,y)$ to the optical flow components $(u,v)$. The network architecture is illustrated in Fig. \ref{fig:NN_arch} and consists of an input layer, an output layer, and several fully-connected hidden layers $L$, each containing $N$ neurons. During the forward pass, each neuron's activation is computed using learned weights and biases, processed through a non-linear activation function $\sigma$ such as \textit{tanh} or \textit{swish}. For our case we use a network with $4$ hidden layers, $128$ neurons per layer and the \textit{swish} activation function. During training, we minimize a loss function $\Lagr(\theta)$ that encapsulates the functional in Eq.~\ref{eq:sensor_OF} by adjusting the network's trainable parameters $\theta$ using the Adam optimizer. The loss function is defined as:
\begin{align}
     \Lagr(\theta) = \Lagr_{D} + \lambda_1 \Lagr_{R} + \lambda_2 \Lagr_{S}
\end{align}
with
\begin{align}
     &\Lagr_{D} =\int_\Omega \lvert I_x u + I_y v + I_t \rvert^2 \\
     &\Lagr_{R} =  \int_\Omega \lvert \nabla u \rvert^2 + \lvert \nabla v \rvert^2 \label{eq:loss_R}\\
     &\mathcal{L}_{S} = \sum_{\mathbf{x}_i \in S} ||\overline{\mathbf{u}}(\mathbf{x}_i) - \overline{\tau}_{w}(\mathbf{x}_i)||^2
    \label{eq:total_loss}
\end{align}
A key advantage of this multi-objective optimization framework is that each loss component can be evaluated on its own spatial grid. This flexibility allows the data and regularization loss terms to be computed across the entire image plane, while the sensor loss is calculated only at the sparse sensor locations. The gradient terms in Eq.~\ref{eq:loss_R} are computed using automatic differentiation (AD), similar to the residual loss calculation in Physics-Informed Neural Networks (PINNs) \citep{Karniadakis2021, Cai2022, Rohlfs2024}.\\
When the weighting parameter $\lambda_2$ is set to zero, our method functionally reduces to the original Horn-Schunck formulation for individual image pairs. However, our neural network implementation provides a significant advantage through its ability to process multiple frames simultaneously during training. The resulting optical flow field naturally emphasizes persistent flow features while suppressing transient artifacts, thereby eliminating the need for explicit frame averaging.\\
When $\lambda_2 \neq 0$, the algorithm incorporates the sparse measurements to create a physically-anchored flow field estimation. This explicit physical grounding mitigates systematic errors without requiring prior knowledge of oil-film properties or pressure gradients, providing a robust foundation for quantitative wall shear stress analysis.
The SENSE algorithm is implemented in Python using the \texttt{Keras 3.6} library with the \texttt{TensorFlow 2.16} backend, with training structured as follows:
\begin{itemize}
    \item Optimization via the Adam algorithm \citep{Kingma2015} using a cosine-decaying learning rate schedule that transitions from an initial value of $10^{-3}$ to $10^{-5}$ over the training duration
    \item Mini-batch processing with 2048 samples per batch for both data ($\Lagr_{D}$) and regularization ($\Lagr_{R}$) terms
    \item Training for 2000 epochs on an Nvidia RTX 4060 Ti GPU, with typical training times of approximately 20 seconds per 100,000 spatial points in the dataset
\end{itemize}

\section{Results}
\label{sec:results}

\subsection{Quantitative Optical Flow without Sensor Information}

To determine a performance baseline for the different optical flow algorithms, we compare their outputs with the reference measurements obtained from MEMS sensors. For this comparison our SENSE algorithm is configured with $\lambda_2$ set to zero, while the regularization weight is set to $\lambda_1 = 10^{-6}$. Apart from the algorithm itself, there are two key parameters that significantly influence the reconstruction accuracy: the number of frames evaluated and the spatial resolution of each frame. As discussed in Sec.\ref{sec:SENSE}, a reliable steady-state solution cannot be derived from a single image pair. However, extended sequences can be compromised by oil film degradation or depletion, which introduces distortions in the estimated flow field \citep{Liu2008}. Similarly, the spatial resolution presents a trade-off: while the linearized optical flow equation is most accurate for small displacements ($O<1$ pixel), excessive downsampling can lead to the loss of critical flow features.\\
\begin{figure}[htbp]
\centering
\includegraphics[width=0.45\textwidth]{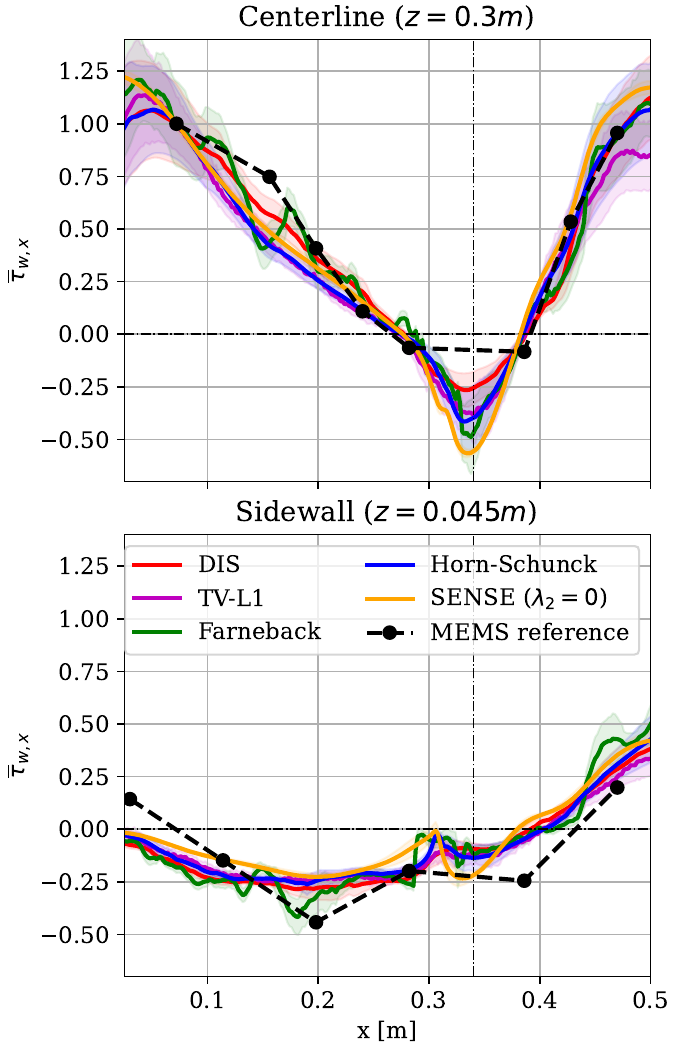}
\caption{Comparison of streamwise shear stress distributions estimated by different optical flow methods against MEMS reference data at $z=0.3$m (top) and $z=0.045$m (bottom). Shaded areas represent standard deviation over 40 frames.}\label{fig:tauw_x}
\end{figure}
In Fig. \ref{fig:tauw_x} we compare the streamwise shear stress distribution from the reference measurements with the averaged OF estimates at two different spanwise positions (see Fig. \ref{fig:SENSE_layout} for the sensor locations relative to the flow features). The results are normalized based on the maximum shear stress measured by the MEMS sensors and the shaded region around the lines highlights the standard deviation of the OF results over a sequence of 40 images. The source images have a size of  $277 \times 300$ pixels translating to a spatial resolution of $2$ mm/px.  which was found to be a value where all algorithms yield very usable results. The centerline distribution ($z = 0.3$ m) exhibits a characteristic decrease in shear stress due to the adverse pressure gradient in the diffuser until the mean flow separates at around $x\approx0.27$ m. Downstream of the ramp ($x>0.34$ m), the flow reattaches at $x\approx 0.39$ m and the shear stress increases again. While all optical flow algorithms capture this general trend, notable differences exist. The smoothest distributions are obtained from the Horn-Schunck and our SENSE algorithm due to their L2 regularization term which penalizes large variations while the Farneback and TV-L1 algorithms tend to have more noise in their outputs when using the default hyper-parameters. The variation over the sequence length is similar for all classical OF approaches and increases towards the frame boundaries where the effect of vanishing oil is most pronounced. The SENSE algorithm has only a very small variation demonstrating the robustness of the multiple-frame training approach. Near the sidewall ($z = 0.045$ m), the flow separates earlier due to the presence of counter-rotating corner vortices, with the reattachment point located further downstream (\cite{Steinfurth2024}). Although the optical flow algorithms qualitatively capture this behavior, the quantitative error relative to the reference measurements increases, and all OF methods tend to predict the separation and reattachment points further upstream compared to the MEMS-implied locations.\\ 
\begin{figure}[htbp]
\centering
\includegraphics[width=0.45\textwidth]{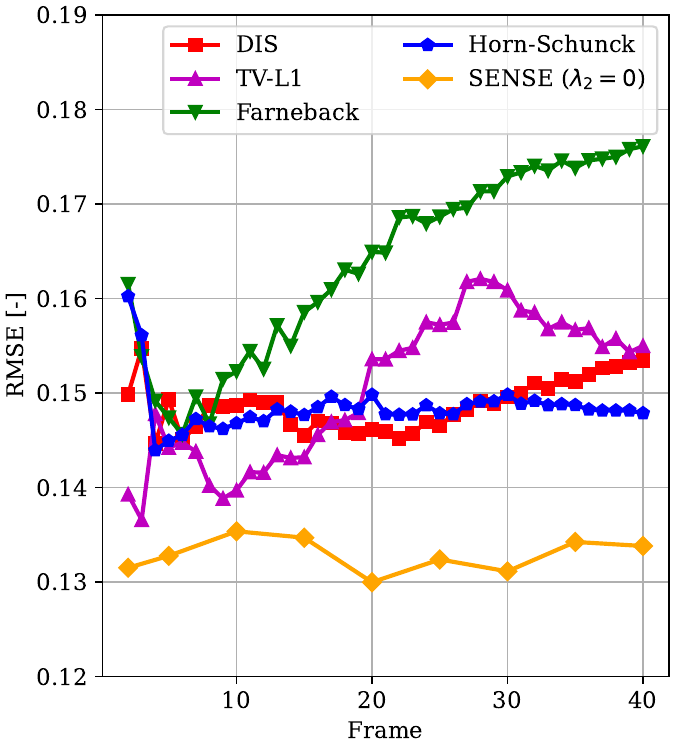}
\caption{RMSE between the reference sensor measurements and the average optical flow estimations for different algorithms and sequence lengths.}\label{fig:RMSE_frame}
\end{figure}

To further quantify the effect of sequence length and the reconstruction error, we compute the root-mean-squared error (RMSE) for both shear stress components between the reference sensor measurements and the average optical flow estimations at all 26 sensor measurement positions. Figure \ref{fig:RMSE_frame} presents the RMSE as a function of sequence length. For $n_{frames} < 10$, all classical algorithms exhibit comparable performance with considerable frame-to-frame variation for very short sequences ($n_{frames} < 5$). For longer sequences, the performance of the Farnebäck and TV-L1 algorithms significantly decreases, while DIS and Horn-Schunck maintain relatively stable RMSE values. The multi-frame prediction capabilities of the SENSE algorithm significantly outperform the averaged snapshot solutions for all sequence lengths with a minimum located at $n_{frames} = 20$. This highlights the ability of SENSE to extract consistent and accurate flow information from longer temporal sequences, mitigating the impact of transient noise and oil film variations.\\
\begin{figure}[htbp]
\centering
\includegraphics[width=0.45\textwidth]{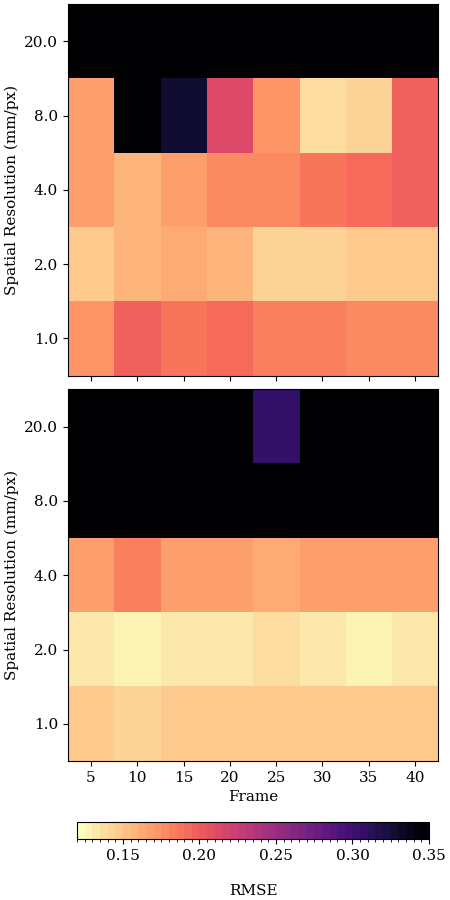}
\caption{RMSE heatmap for the computed shear stress for different spatial resolutions and sequence lengths; Top: Horn-Schunck algorithm; Bottom: SENSE without sensor information $\lambda_2=0$}\label{fig:RMSE_heatmap}
\end{figure}
In order to investigate the effect of the spatial resolution, we compute the RMSE for 6 different image scales ranging from $20$ mm/px to $1.0$ mm/px across varying sequence lengths. The resulting heat maps, displayed in Fig. \ref{fig:RMSE_heatmap}, compare the Horn-Schunck algorithm as the most comparable classical OF method on the left, with our SENSE algorithm (with $\lambda_2=0$) on the right. As previously mentioned, $2$ mm/px appears to be optimal for this experimental setup, yielding the lowest RMSE values for both Horn-Schunck and SENSE ($\lambda_2=0$). At $4$ mm/px the accuracy is worse, but remains acceptable with a maximum displacement over the entire image sequence of approximately $1.5$ px in this case. At lower resolutions ($\geq 8$ mm/px), the reconstruction quality deteriorates significantly as the maximum displacements increase and the available image information decreases. At higher spatial resolutions, both methods exhibit increased errors, although the Horn-Schunck algorithm experiences a substantially larger increase. The RMSE for the SENSE approach remains comparable to the lowest error achieved by the Horn-Schunck, suggesting a greater resilience to noise and small-scale variations in the flow field.

\subsection{Multi-Sensor Improvements}
\begin{figure}[htbp]
\centering
\includegraphics[width=0.45\textwidth]{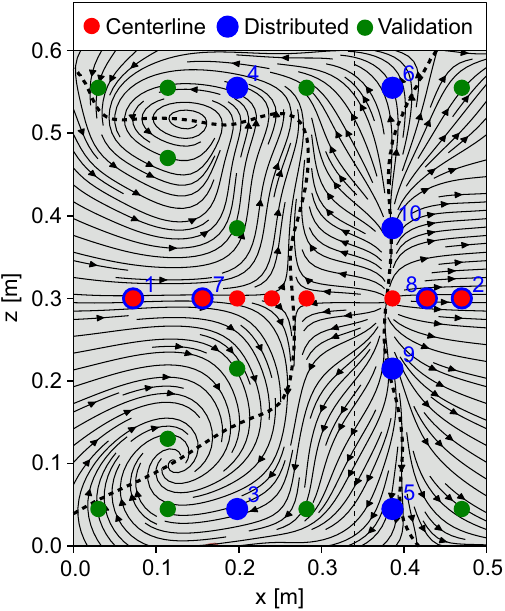}
\caption{Top view of the diffuser with streamlines indicating the near-wall topology obtained with baseline SENSE. Sensor positions are highlighted for reference. Red: Sensors used for the centerline strategy; Blue: Sensors used for the distributed strategy; Green: Sensors used for validation}\label{fig:SENSE_layout}
\end{figure}

Having established the baseline performance of the SENSE algorithm without active sensor feedback ($\lambda_2 = 0$), we now examine the impact of incorporating sparse sensor measurements on the accuracy of wall shear stress estimation. For simplicity, we consider two different sensor placement strategies. The first approach utilizes only the eight sensors located along the centerline of the test section, representing a minimum effort setup that only requires an oil-film vizualisation and a few measurements in a straight line without prior knowledge of the flow. These sensors are highlighted in red in Fig. \ref{fig:SENSE_layout}. The second approach strategically positions up to ten sensors in regions of high shear stress magnitude (blue dots). In a practical wind tunnel experiment with unknown flow characteristics, this approach can be implemented by initially performing an oil-film visualization, analyzing the images in-situ with a classical optical flow method (e.g. using TUBflow \citep{Rohlfs2024a}), and then placing sensors in areas exhibiting high optical flow magnitudes.\\
\begin{figure}[htbp]
\centering
\includegraphics[width=0.49\textwidth]{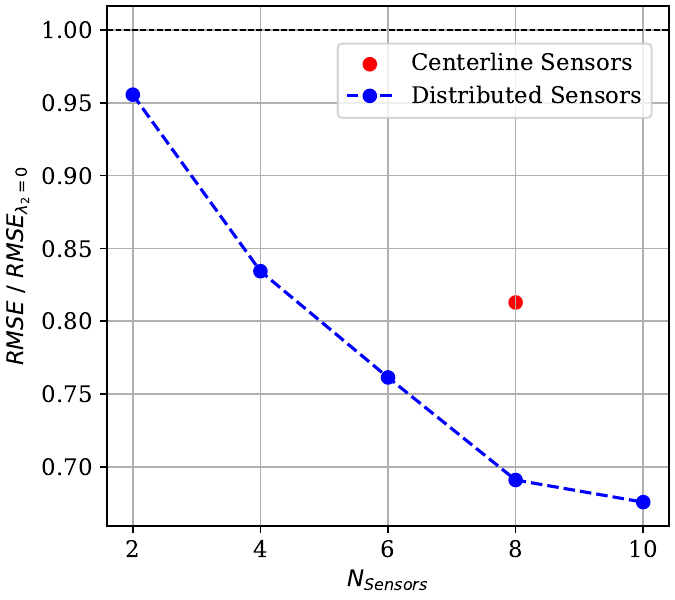}
\caption{Error reduction relative to the baseline algorithm ($\lambda_2=0$) due to additional sensors in the training dataset}\label{fig:SENSE_RMSE}
\end{figure}

Figure \ref{fig:SENSE_RMSE} displays the achievable error reduction for the shear stress magnitude relative to the baseline case without any sensors ($\lambda_2=0$) for the different sensor placement strategies. To ensure a fair comparison, the RMSE is computed only at positions not included in the training dataset (green dots in Fig. \ref{fig:SENSE_layout}). When using the 8 centerline sensors in the training database, the RMSE can be reduced by $\approx20\%$. However, if the same number of sensors are distributed also in the spanwise direction the RMSE can be reduced by over $30\%$. Notably, a $15-20\%$ reduction can be achieved with only four strategically placed sensors - two at the entrance and exit of the diffuser, and one each towards the sidewall where the reverse flow is most pronounced.\\
\begin{figure}[htbp]
\centering
\includegraphics[width=0.45\textwidth]{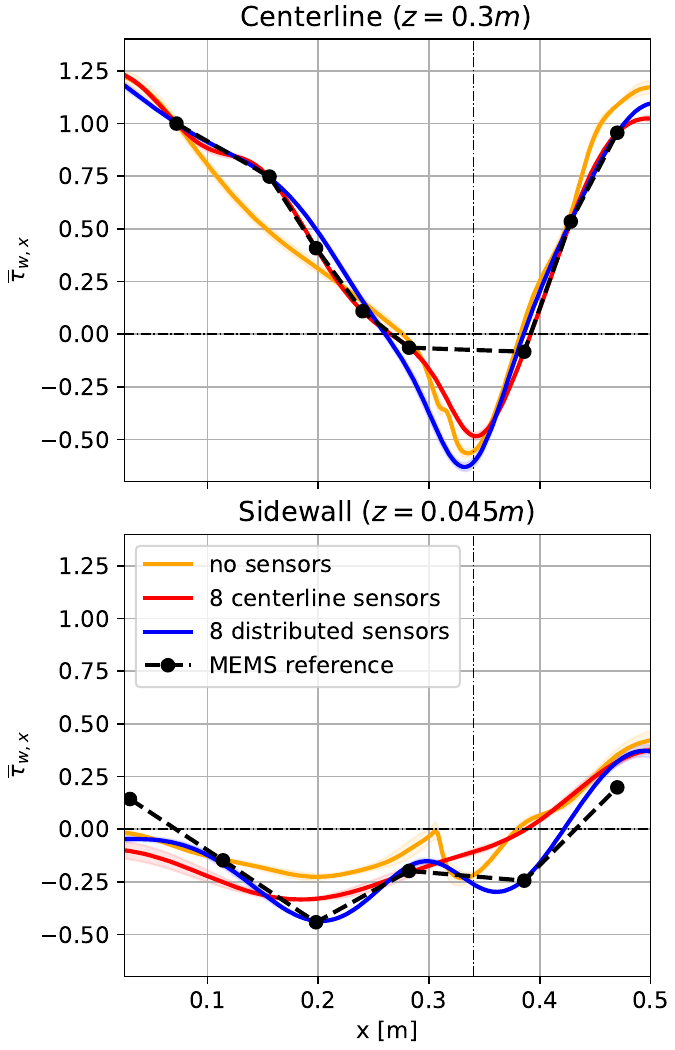}
\caption{Sensor effect on the streamwise shear stress distributions at $z=0.3$m (top) and $z=0.045$m (bottom).}\label{fig:SENSE_MEMS}
\end{figure}
In Fig. \ref{fig:SENSE_MEMS} we compare the streamwise shear stress distributions estimated by the SENSE algorithm with and without sensor information. The source images are the same as for Fig. \ref{fig:tauw_x} with a sequence length of 40 images and a spatial resolution of $2$ mm/px. Along the centerline, the SENSE algorithm trained with the centerline sensors closely replicates the reference shear stress profile. This is expected, as all reference points along this line were included in the training data. The configuration using the distributed sensors also achieves an improved accuracy compared to the baseline, even with fewer sensors directly on the centerline. Near the sidewall the benefits of the distributed sensor placement become much more apparent with its estimation closely tracking the MEMS reference data. Notably, the centerline configuration yields a modest improvement over the baseline near the sidewall, demonstrating that the sensor information exerts a global influence, regularizing the solution across the entire domain, not just locally around the sensors.

\begin{figure*}[htbp]
    \centering
    \includegraphics[width=0.9\textwidth]{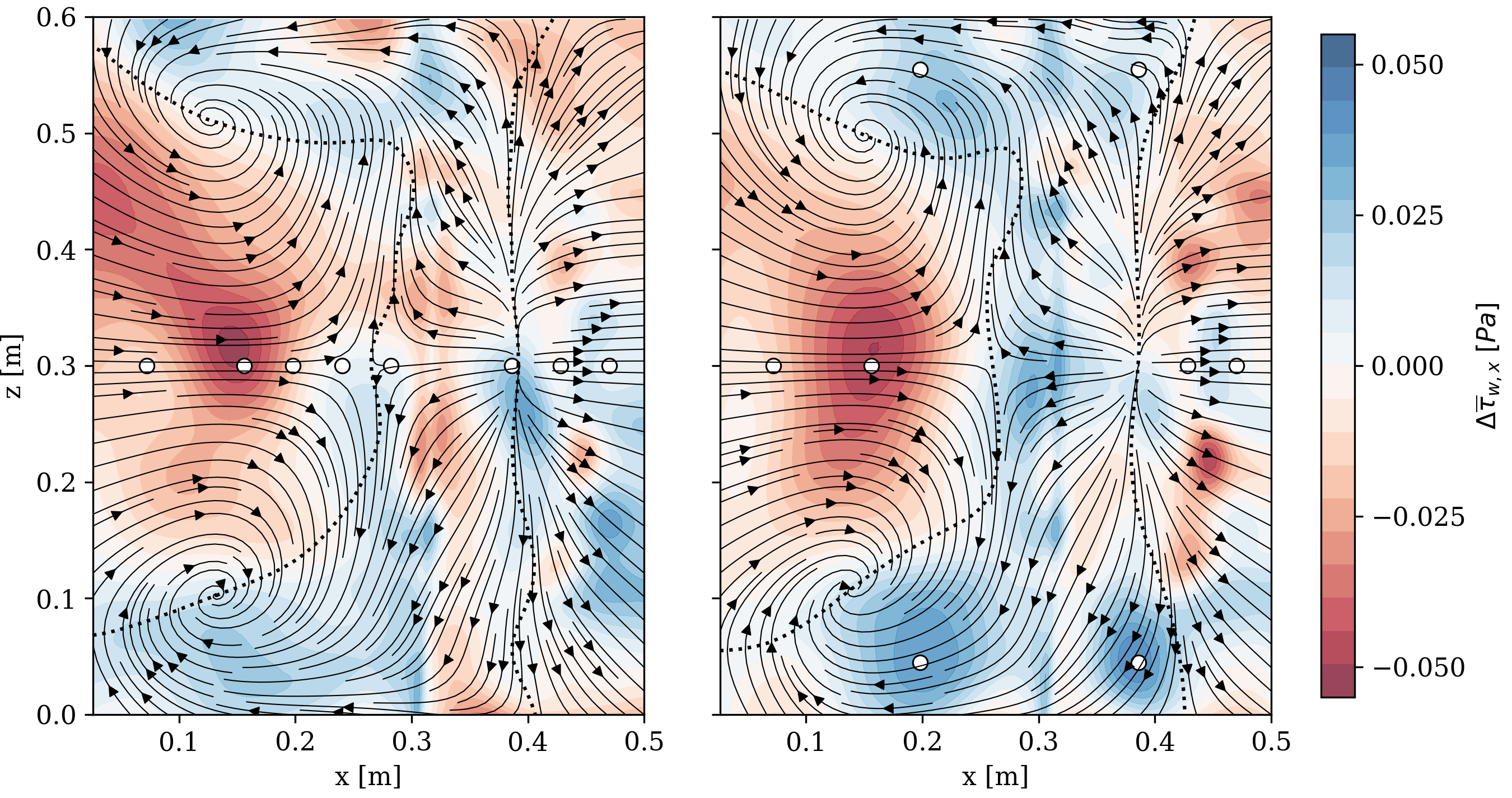}
    \caption{Spatial effect of the sensor regularization. Streamlines indicate the near-wall topology, while contours depict the difference in $\tau_{w,x}$ relative to the baseline. Sensor locations are marked with white dots. Left: centerline sensors, right: distributed sensors }\label{fig:SENSE_contour}
\end{figure*}

Fig. \ref{fig:SENSE_contour} further explores the impact of the sensor regularization across the entire domain. The streamlines show the computed shear stress topology for both the centerline (left panel) and distributed (right panel) sensor configurations, while contours represent the difference in streamwise shear stress ($\tau_{w,x}$) compared to the baseline SENSE result ($\lambda_2=0$). Both sensor-constrained configurations yield qualitatively similar flow topologies that effectively capture the primary three-dimensional flow structures, including the main separation and reattachment lines. The contour plots highlight localized regions where the sensor-constrained solutions deviate significantly from the baseline. The largest positive difference (indicating an increase in estimated $\tau_{w,x}$ relative to the baseline) for both configurations is observed near the second centerline sensor. This corresponds to an area where the baseline SENSE and classical optical flow methods were shown to underestimate the shear stress magnitude compared to the MEMS reference (compare Fig \ref{fig:tauw_x}). Similarly, for the distributed sensor strategy (right panel), notable deviations are evident near the sidewall sensors, consistent with the improved accuracy demonstrated in Fig. \ref{fig:SENSE_MEMS}. Apart from these differences that can be clearly linked to the respective sensor measurements in the training dataset, there are many other regions with modifications to the estimated shear stress fields. This further underscores the global regularizing effect of the sensor loss term ($\mathcal{L}_S$), demonstrating that incorporating sparse measurements influences the predicted flow field beyond the specific measurement points.

\section{Conclusions and Future Work}
\label{sec:conclusion}
This study introduced SENSE (Sensor-Enhanced Neural Shear Stress Estimation), a novel data-driven methodology designed to enhance the quantitative accuracy of wall shear stress ($\tau_w$) fields derived from surface oil-film visualizations. SENSE leverages a Multi-Layer Perceptron neural network to compute the optical flow from image sequences, integrating sparse, high-fidelity shear stress measurements from MEMS sensors directly into the neural network's training via a multi-objective loss function. This approach demonstrated several key advantages over classical optical flow methods. First, by processing multiple frames simultaneously, SENSE effectively mitigates temporal noise and variations inherent in the oil-film evolution without requiring explicit averaging, proving more robust especially for longer image sequences. Second, the inclusion of a sensor-based loss term anchors the optical flow solution to physical ground-truth values. This significantly improves quantitative accuracy, achieving over 30\% reduction in RMSE on validation sensors when employing a strategically distributed sensor layout. The results indicate that even a small number of sensors placed in regions identified as significant (e.g. areas exhibiting high optical flow magnitudes) can substantially improve the overall field accuracy. Furthermore, the sensor data was observed to provide a global regularization effect that reaches far beyond the immediate vicinity of the sensor locations.\\
As such, the proposed method is a promising tool to elevate oil-film visualizations from a primarily qualitative tool to a reliable quantitative measurement technique. For future studies, applying SENSE to a wider range of flow regimes and geometries (e.g. compressible flows and curved surfaces) would validate its versatility. Additionally, other regularization terms within the loss function (e.g., physics-based constraints if applicable) or different neural network architectures could be explored to further improve the accuracy. Finally the confidence in the results could be enhanced by adapting the SENSE framework to provide not just the $\tau_w$ field but also a spatially-resolved map of the prediction uncertainty using Bayesian neural networks or ensemble methods.

\backmatter

\bibliography{references}

\end{document}